# Time-frequency optical filtering: efficiency vs. temporal-mode discrimination in incoherent and coherent implementations

## Michael G. Raymer,[1,2] and Konrad Banaszek[3,4,5]

[1]Department of Physics, Center for Optical, Molecular and Quantum Science, University of Oregon, Eugene, Oregon 97403, USA
[3] Centre for Quantum Optical Technologies, Centre for New Technologies, University of Warsaw, Banacha 2c, 02-097 Warszawa, Poland
[4] Faculty of Physics, University of Warsaw, Pasteura 5, 02-093 Warszawa, Poland
[2] raymer@uoregon.edu
[5] k.banaszek@cent.uw.edu.pl

**Abstract:** Time-frequency (TF) filtering of analog signals has played a crucial role in the development of radio-frequency communications, and is currently being recognized as an essential capability for communications, both classical and quantum, in the optical frequency domain. How best to design optical time-frequency (TF) filters to pass a targeted temporal mode (TM), and to reject background (noise) photons in the TF detection window? The solution for 'coherent' TF filtering is known—the quantum pulse gate—whereas the conventional, more common method is implemented by a sequence of incoherent spectral filtering and temporal gating operations. To compare these two methods, we derive a general formalism for two-stage incoherent time-frequency filtering, finding expressions for signal pulse transmission efficiency, and for the ability to discriminate TMs, which allows the blocking of unwanted background light. We derive the tradeoff between efficiency and TM discrimination ability, and find a remarkably concise relation between these two quantities and the time-bandwidth product of the combined filters. We apply the formalism to two examples—rectangular filters or Gaussian filters—both of which have known orthogonal-function decompositions. The formalism can be applied to any state of light occupying the input temporal mode, e.g., 'classical' coherent-state signals or pulsed single-photon states of light. In contrast to the radio-frequency domain, where coherent detection is standard and one can use coherent matched filtering to reject noise, in the optical domain direct detection is optimal in a number of scenarios where the signal flux is extremely small. Our analysis shows how the insertion loss and SNR change when one uses incoherent optical filters to reject background noise, followed by direct detection, e.g. photon counting. We point out implications in classical and quantum optical communications. As an example, we study quantum key distribution, wherein strong rejection of background noise is necessary to maintain a high quality of entanglement, while high signal transmission is needed to ensure a useful key generation rate.

## 1. Introduction

Time-frequency (TF) filtering of optical communication signals is playing an increasingly important role as quantum communication or quantum-limited classical communication systems are being more frequently deployed. Such filtering of optical signals is important for rejecting background noise in optical communication systems, for example photon-counted free-space optical links such as Earth-orbit to ground stations. [1, 2] Noise rejection becomes especially critical for quantum communication, where the distribution of quantum entanglement across large distances is paramount. When background light is present in free-space quantum optical links, it can create false counts in detectors, reducing the quality of distributed entanglement and thus the key rate. [3] Another venue where optimal TF filtering is crucial is deep-space communications, where detected photons are few and far between (the so-called photon-starved regime). As reviewed in [1], for this case it is known that near optimality can be achieved by transmitting information encoded in the temporal location of coherent (e.g. laser) pulses (e.g. pulse-position modulation), and receiving the signal using TF filtering followed by direct photon-counting detection. In the photon-starved regime, such a scheme outperforms coherent homodyne or heterodyne detection because it obviates the need for a local oscillator at the detector, which is always accompanied by its inherent shot noise [4].

The general question arises: What is the optimal design for an optical time-frequency filter, with respect to a particular figure of merit for a given application? The answer will depend on understanding what types of TF filters are achievable in the optical domain, and what inherent tradeoffs they operate under. A good TF filter should have a high ability to discriminate between a temporal mode that carries a signal and all other modes that carry only unwanted noise. It should also have high efficiency for passing the desired, targeted temporal mode. In this paper we explore the trade-off between efficiency and mode-discrimination ability of these two types of optical TF filters. Our analysis shows how the insertion loss and SNR change when one uses incoherent optical filters to reject background noise, followed by direct detection, e.g. photon counting.

Optical TF filters fall into two categories: incoherent and coherent. Incoherent TF filters (ITFF) act only on the intensity, and not the phase, of light in either the time or frequency domain, while coherent TF filters (CTFF) act in a manner that depends on the complex field amplitude and thus on its temporal or spectral phase structure. A common realization of a TF filter is a sequence of an incoherent spectral filter (e.g., a spectrometer) followed by an incoherent temporal filter (e.g., a shutter), as shown in Fig. 1.

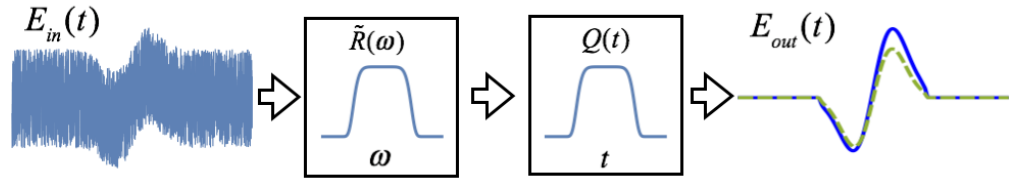

**Figure 1.** A sequence of two incoherent filters. A (time-stationary) incoherent spectral filter acts first and a (frequency-neutral) incoherent temporal filter second. The input consists of the desired signal $A_0 f(t)$ and a noise field $y(t)$. The output (simulated, solid curve) is a spectrally and temporally truncated field, which contains mostly the desired signal and some residual smoothed noise. The dashed curve shows the signal expected if no noise were present.

In contrast, a coherent TF filter acts like a time-non-stationary spectral filter or equivalently a frequency-non-neutral temporal filter. CTFFs can be implemented straightforwardly in the radio domain for classical communication systems, where fast electronics can process voltages directly; this enables, for example, code-division multiple access (CDMA), used in many mobile phone systems. In the optical domain the carrier frequencies are too high to operate upon directly, and, until the recent invention of the 'quantum pulse gate,' no device was known that could efficiently demultiplex optical pulses according to their orthogonal, overlapping temporal shapes [5, 6, 7, 8, 9, 10] The quantum pulse gate (QPG) functions equally well with classical or quantum signals, as it operates as a TF mode filter or demultiplexer, in principle without loss, leaving intact the state of the field, which may be classical (coherent state, thermal state, etc.) or quantum (single-photon, squeezed, etc.) The name QPG arose from its importance in quantum communication research as reviewed in [11]. For example, enabling the use of higher-dimensional encoding schemes, which facilitates increased information capacity per photon and increased security in quantum communication [12, 13, 14] when compared to two-dimensional encoding. And for classical signals in the photon-starved regime, it has been argued that CTFFs provides the highest possible information content per photon, and thus the highest photon information efficiency (PIE). [2]

The quantum pulse gate (QPG) uses pulsed nonlinear optical sum-frequency generation to operate on complex optical field amplitudes. [5, 9] The weak incoming signal with the central frequency $\omega_S$ enters a nonlinear optical crystal in the presence of a strong laser pulse with the central frequency $\omega_P$, which up-converts the signal to frequency $\omega_S + \omega_P$. Filtering takes place because only the portion of the signal (or noise) whose temporal waveform matches the laser pulse is converted. One experimental study to date has used this method to demonstrate improved noise rejection of a QPG compared to an ITFF. [15] There are two other proposed methods for implementing an optical CTFF. One is by sum-frequency generation in an optical cavity. [16] The other is by rapidly varying the input coupling to an optical cavity. [17] Here we simply assume that CTFFs are available, and compare their operation to ITFFs. In addition, 'hybrid' methods have also been proposed and in-part implemented, using electrooptic modulators in combination with storage loops for projective measurement. [18]

Free-space communication systems use light pulses in a beam-like geometry, and fiber-optic systems use guided spatial modes. In such cases, the pulsed field can be represented by an expansion in a discrete set of temporal modes (TMs), which are pulse-like solutions of the wave equation in the medium. [18, 11, 19, 20] In the following we typically take the information-carrying signal to occupy only one of the TMs (whose form is to be determined or simply chosen) and the noise is assumed to occupy equally all TMs. An ideal TF filter would block all TMs except the one carrying the signal, which of course also carries with it 'one mode worth' of the background noise. The form of a TM is specified by its complex spectral amplitude $\tilde{E}(\omega)$, which also determines its spatial propagation. Because the propagation is 'trivial,' in this paper we give the form of the field at an arbitrary longitudinal spatial point ($z = 0$).

## 2. General TF filtering

The seminal works of David Slepian and coworkers considered essential questions such as what pulsed radio-frequency wave forms have their energy maximally concentrated in both time and frequency. [21,22] And it has been known for decades that radio-frequency TF filters, also called linear time-varying filters [ 23 ], can efficiently discriminate, separate, or demultiplex, a single temporal mode (TM) from a time-frequency continuum. Here we assume such operations are available in the optical regime and discuss implementation later.

The most general non-amplifying linear filter acts on an input signal $E_{in}(t)$ (positive-frequency part or analytic signal) to produce an output signal $E_{out}(t)$ as:

$$E_{in}(t) \rightarrow E_{out}(t) = \int dt' F(t,t') E_{in}(t') \qquad (1)$$

Here and in the following, integrals are taken from $-\infty$ to $+\infty$ unless otherwise stated.

The filter kernel $F(t,t')$ equals zero for $t' > t$ to ensure causality. Note, for example, that for a time-stationary frequency filter, such as an interference bandpass filter, the filter kernel is of the form $F(t-t')$; that is *not* an example of a time-varying filter.

In the frequency domain, Eq.(1) becomes:

$$\tilde{E}_{out}(\omega) = \int d\omega' \, \tilde{F}(\omega,\omega') \tilde{E}_{in}(\omega') \qquad (2)$$

where

$$\tilde{F}(\omega,\omega') = \int dt\, e^{i\omega t} \int dt'\, e^{-i\omega' t'} F(t,t') \tag{3}$$

We use the notation $\mathchar'26\mkern-9mu d\omega \equiv d\omega/2\pi$, which avoids writing $2\pi$ in many places, and reminds us that $\omega/2\pi$ is frequency in Hz. And we use the overtilde to indicate the Fourier transform:

$$\tilde{f}(\omega) = \int dt\, e^{i\omega t} f(t) \;,\;\; f(t) = \int \mathchar'26\mkern-9mu d\omega\, e^{-i\omega t} \tilde{f}(\omega) \tag{4}$$

Any physical kernel can be represented by a Schmidt decomposition (the infinite-dimensional analog of the singular value decomposition), that is a sum over weighted products of basis functions [24]:

$$F(t,t') = \sum_{n=0}^{\infty} \lambda_n \psi_n(t) \varphi^*_{\,n}(t')$$

$$\tilde{F}(\omega,\omega') = \sum_{n=0}^{\infty} \lambda_n \tilde{\psi}_n(\omega) \tilde{\varphi}^*_n(\omega') \tag{5}$$

where again the overtilde indicates the Fourier transform. The non-negative, real singular values $\lambda_n$ are determined by the allowed solutions of the integral equation (expressed in two equivalent forms):

$$\int dt'\, F(t,t') \varphi_n(t') = \lambda_n \psi_n(t)$$

$$\int \mathchar'26\mkern-9mu d\omega'\, \tilde{F}(\omega,\omega') \tilde{\varphi}_n(\omega') = \lambda_n \tilde{\psi}_n(\omega) \tag{6}$$

Every $\lambda_n$ lies between zero and one, with a constraint on the sum of their squares:

$$\sum_{n=0}^{\infty} \lambda_n^{\,2} = \int \mathchar'26\mkern-9mu d\omega \int \mathchar'26\mkern-9mu d\omega' \,|\, \tilde{F}(\omega,\omega')\,|^2 \tag{7}$$

which may be greater than one. We order the $\lambda_n$'s in decreasing order, and take $\lambda_0$ to be the largest. Orthonormality of the basis (mode) functions is expressed by:

$$\int dt\, \psi^*_{\,m}(t) \psi_n(t) = \delta_{mn}$$

$$\int dt\, \varphi^*_{\,m}(t) \varphi_n(t) = \delta_{mn} \tag{8}$$

And, likewise, in the frequency domain:

$$\int \mathchar'26\mkern-9mu d\omega\, \tilde{\psi}^*_m(\omega) \tilde{\psi}_n(\omega) = \delta_{mn}$$

$$\int \mathchar'26\mkern-9mu d\omega\, \tilde{\varphi}^*_m(\omega) \tilde{\varphi}_n(\omega) = \delta_{mn} \tag{9}$$

The expansion into temporal modes leads to:

$$E_{out}(t) = \sum_{n=0}^{\infty} \lambda_n \psi_n(t) \int dt' \varphi_n^*(t') E_{in}(t') \tag{10}$$

the interpretation being that the input pulse is projected onto the set of input functions $\varphi_n(t')$, each of which is paired with a unique output function $\psi_n(t)$. Thus, if the input pulse equals a particular input basis function, say $\varphi_N(t')$, then the output pulse equals $\lambda_N \psi_N(t)$. The 'perfect' TF filter, for our purposes, would be one having one of the singular values, say $\lambda_0$, equal to 1, with all others zero. In that case, the kernel is *separable (factorable)*: $F(t,t') = \lambda_0 \psi_0(t) \varphi_0^*(t')$. Then Eq.(10) becomes:

$$E_{out}(t) = \lambda_0 \psi_0(t) \int dt' \varphi_0^*(t') E_{in}(t') \tag{11}$$

### 3. Mode discrimination-efficiency tradeoff

The goal of a TF filter is to have both good TM discrimination ability (to block unwanted noise) and high efficiency (to transmit the wanted signal pulse). *Efficiency* is defined as the squared singular value for the wanted 'target' TM, i.e. that TM having the highest transmission: $\eta = \lambda_0^2$. *Descrimitivity* is defined as its ratio to the sum of transmission for all other TMs:

$$\xi = \frac{\lambda_0^2}{\sum_{n=0}^{\infty} \lambda_n^2} \tag{12}$$

An ideal TF filter will have both $\eta$ and $\xi$ close to one. This is achieved only if $\lambda_0 \cong 1$ and $\lambda_{n \neq 0} \cong 0$ .

For incoherent TF filters there exists a generic tradeoff between descrimitivity and efficiency. Generally, when $\eta$ goes to one, $\xi$ goes to zero; and vice versa. Below we calculate and plot this tradeoff as a function of operational parameters for a few types of TF filters. Thereby we quantify the superior performance of coherent filters, which do not suffer the same tradeoff, over incoherent filters.

Note, there is no consensus on naming $\xi$ in the literature. It has been called *separability* [9, 8] and, in a modified form, *selectivity* [25]. To avoid ambiguity, we introduce the name *descrimitivity* (the ability to discriminate), for which there is precedent in other technical literature. We continue to use the term *selectivity* for the product $\eta\xi$, as in [8\, 9, 11].

### 4. Background noise rejection

When a wanted signal is embedded in a noise background, an ideal TF filter would completely transmit the signal and reject as much noise as possible. In principle, it could reject all the noise except that which occupies the signal mode. Model the input signal as a sum of a targeted signal pulse $f(t)$ plus a white-noise background $y(t)$:

$$E_{in}(t) = A_0 f(t) + y(t) \quad , \quad \tilde{E}_{in}(\omega) = A_0 \tilde{f}(\omega) + \tilde{y}(\omega) \tag{13}$$

where $|A_0|^2$ is the energy (or mean photon number) in the signal pulse, normalized according to $\int |f(t)|^2 \, dt = \int |\tilde{f}(\omega)|^2 \, d\omega = 1$. The noise has correlation function $\langle y^*(t) y(t') \rangle = N_y \delta(t-t')$, and in the frequency domain $\langle \tilde{y}^*(\omega) \tilde{y}(\omega') \rangle = N_y 2\pi \delta(\omega - \omega')$, where $N_y$ is noise power spectral density (noise energy per unit time-bandwidth area). The detector integrates flux over the time of the received pulse:

$$
\begin{aligned}
W &= \int dt \, | E_{out}(t) |^2 = \int d\omega \, | \tilde{E}_{out}(\omega) |^2 \\
&= \int d\omega' \int d\omega'' G(\omega', \omega'') \tilde{E}_{in}^{\,*}(\omega') \tilde{E}_{in}(\omega'')
\end{aligned}
\tag{14}
$$

where

$$
\begin{aligned}
G(\omega', \omega'') &= \int d\omega \, \tilde{F}^*(\omega, \omega') \cdot \tilde{F}(\omega, \omega'') \\
&= \sum_{n=0}^{\infty} \lambda_n^{\,2} \tilde{\varphi}_n(\omega') \tilde{\varphi}_n^{\,*}(\omega'')
\end{aligned}
\tag{15}
$$

It is useful to write the signal and noise energies separately, $W = W_f + W_y$, where:

$$
\begin{aligned}
W_f &= |A_0|^2 \sum_{n=0}^{\infty} \lambda_n^{\,2} \left| \int d\omega \, \tilde{\varphi}_n^{\,*}(\omega) \tilde{f}(\omega) \right|^2 \\
&\rightarrow |A_0|^2 \lambda_0^{\,2} \, (ideal)
\end{aligned}
\tag{16}
$$

and:

$$
\begin{aligned}
\langle W_y \rangle &= \int d\omega' \int d\omega'' G(\omega', \omega'') \langle \tilde{y}^*(\omega') \tilde{y}(\omega'') \rangle \\
&= N_y \sum_{n=0}^{\infty} \lambda_n^{\,2} = N_y \eta / \xi
\end{aligned}
\tag{17}
$$

where the signal-noise cross terms are zero, and where $\xi$ is the descrimitivity (mode discrimination ability) given in Eq.(12). We see that the ratio of signal to noise is:

$$SNR = \frac{W_f}{\langle W_y \rangle} = \frac{|A_0|^2}{N_y} \xi \tag{18}$$

Thus, in this context the descrimitivity characterizes noise rejection. A perfect TF filter, which can be approached by a QPG, will have $\xi = 1$ and thus the optimal achievable SNR equals $|A_0|^2 / N_y$.

## 5. Incoherent TF filters

Whether a filter is coherent or incoherent is determined by the form of $F(t, t')$. An *incoherent frequency-domain filter* is described by $\tilde{F}_F(\omega, \omega') = 2\pi\delta(\omega - \omega')\tilde{R}(\omega')$, so that

$$\tilde{E}_{out}(\omega) = \tilde{R}(\omega)\tilde{E}_{in}(\omega) \tag{19}$$

That is, the spectrum is simply multiplied by a windowing function, e.g., a spectral bandpass filter. Then the transmitted pulse energy is independent of the spectral phase structure of the input:

$$W_F = \int |\tilde{E}_{out}(\omega)|^2 \, d\omega = \int |\tilde{R}(\omega)|^2 |\tilde{E}_{in}(\omega)|^2 \, d\omega \tag{20}$$

and so is seen to be incoherent. This filter is 'stationary' in the time domain, $F_F(t, t') = R(t - t')$, and acts as:

$$E_{out}(t) = \int dt' R(t - t') E_{in}(t') \tag{21}$$

where the filter's response function satisfies $R(t - t') = 0$ for $t' > t$.

An *incoherent time-domain filter* is described by $F_T(t, t') = \delta(t - t')Q(t')$, so that

$$E_{out}(t) = Q(t)E_{in}(t) \tag{22}$$

That is, the field is simply multiplied by a time-dependent windowing function. Then the transmitted pulse energy is independent of the temporal phase structure of the input:

$$W_T = \int |E_{out}(t)|^2 \, dt = \int |Q(t)|^2 |E_{in}(t)|^2 \, dt \tag{23}$$

and so the filter is incoherent. This filter is 'stationary' in the frequency domain, $\tilde{F}_T(\omega, \omega') = \tilde{Q}(\omega - \omega')$, and acts as:

$$\tilde{E}_{out}(\omega) = \int d\omega' \tilde{Q}(\omega - \omega')\tilde{E}_{in}(\omega') \tag{24}$$

A lossless coherent filter can have an arbitrary form as long as it satisfies unitarity when acting on a complete of TMs; therefore, the transmitted pulse energy is most generally:

$$W_{coh} = \int |E_{out}(t)|^2\, dt$$
$$= \int dt'' \int dt'\, G(t',t'') E^*_{in}(t'') E_{in}(t') \tag{25}$$

where

$$G(t',t'') = \int dt\, F^*(t,t'') F(t,t') \tag{26}$$

which *does* depend on the phase structure of the input. In the frequency domain the transmitted energy is:

$$W_{coh} = \int d\omega \int d\omega'\, \tilde{G}(\omega,\omega') \tilde{E}^*_{in}(\omega') \tilde{E}_{in}(\omega'') \tag{27}$$

where

$$\tilde{G}(\omega,\omega') = \int d\omega\, \tilde{F}^*(\omega,\omega') \tilde{F}(\omega,\omega'') \tag{28}$$

Perhaps the most common type of TF filter is what we call a *sequential incoherent filter* (SIF), consisting of a sequence of two incoherent filters—one time-domain and one frequency-domain—in either order. Consider a SIF with the frequency filter placed first and time filter second:

$$E_{out}(t) = \int dt'\, F_T(t,t') \int dt''\, F_F(t',t'') E_{in}(t'')$$
$$= \int dt''\, F_{TF}(t,t'') E_{in}(t'') \tag{29}$$

where

$$F_{TF}(t,t'') = \int dt'\, F_T(t,t') F_F(t',t'')$$
$$= \int dt'\, \delta(t-t') Q(t') R(t'-t'') \tag{30}$$
$$= Q(t) R(t-t'')$$

One reason to consider this order is that temporal gating can be done at the electronics level, after the photodetector. In the frequency domain this is written:

$$\tilde{E}_{out}(\omega) = \int d\omega'\, \tilde{F}_T(\omega,\omega') \int d\omega''\, \tilde{F}_F(\omega',\omega'') \tilde{E}_{in}(\omega'')$$
$$= \int d\omega''\, \tilde{F}_{TF}(\omega,\omega'') \tilde{E}_{in}(\omega'') \tag{31}$$

where

$$\tilde{F}_{TF}(\omega,\omega'') = \int d\omega'\, \tilde{F}_T(\omega,\omega') \tilde{F}_F(\omega',\omega'')$$
$$= \int d\omega'\, \tilde{Q}(\omega-\omega') \delta(\omega'-\omega'') \tilde{R}(\omega')$$
$$= \tilde{Q}(\omega-\omega'') \tilde{R}(\omega'') \tag{32}$$

If the time and frequency filters are swapped, with the time filter acting first, the results are:

$$E_{out}(t) = \int dt' F_F(t,t') \int dt'' F_T(t',t'') E_{in}(t'')$$
$$= \int dt'' F_{FT}(t,t'') E_{in}(t'') \tag{33}$$

where

$$F_{FT}(t,t'') = \int dt' F_F(t,t') F_T(t',t'')$$
$$= \int dt' R(t-t') \delta(t'-t'') Q(t')$$
$$= Q(t'') R(t-t'') \tag{34}$$

and in the frequency domain:

$$\tilde{E}_{out}(\omega) = \int d\omega' \tilde{F}_F(\omega,\omega') \int d\omega'' \tilde{F}_T(\omega',\omega'') \tilde{E}_{in}(\omega'')$$
$$= \int d\omega'' \tilde{F}_{FT}(\omega,\omega'') \tilde{E}_{in}(\omega'') \tag{35}$$

where

$$\tilde{F}_{FT}(\omega,\omega'') = \int d\omega' \tilde{F}_F(\omega,\omega') \tilde{F}_T(\omega',\omega'')$$
$$= \int d\omega' \delta(\omega-\omega') \tilde{R}(\omega') \tilde{Q}(\omega'-\omega'')$$
$$= \tilde{R}(\omega) \tilde{Q}(\omega-\omega'') \tag{36}$$

A symmetry exists between these filter orderings. If in the frequency domain the time filter is Hermitian, $\tilde{Q}^*(\omega'-\omega) = \tilde{Q}(\omega-\omega')$, and the frequency filter is real, $\tilde{R}^*(\omega) = \tilde{R}(\omega)$, then it is not hard to show that the singular values are independent of the filter order, and the forms of the input and output modes are simply swapped. (The same holds true if $\tilde{Q}(\omega-\omega')$ is Hermitian in the time domain and $R(t)$ is real in the time domain.)

Although a SIF is comprised of two (or more) incoherent filters, its overall affect is that of a coherent filter, but with restricted properties. This can be seen by calculating the transmitted energy, for example:

$$W_{TF} = \int |E_{out}(t)|^2 \, dt = \int dt'' \int dt' G_{TF}(t',t'') E^*_{in}(t'') E_{in}(t') \tag{37}$$

where

$$G_{TF}(t',t'') = \int dt F_{TF}^*(t,t'') F_{TF}(t,t')$$
$$= \int dt |Q(t)|^2 \, R^*(t-t'') R(t-t') \tag{38}$$

which does depend on the signal's phase structure. Comparing this result to that in Eq.(25), we see that $W_{TF}$ is a subset or restricted case of $W_{coh}$.

This result agrees with the idea that we can create a nearly coherent pulse by applying a SIF to a broadband incoherent source of light, such as a blackbody emitter. If $y(t)$ denotes a stationary, white-noise field with power $N_y$ and correlation function $\left\langle y^*(t') y(t'') \right\rangle = N_y \delta(t'-t'')$, then the transmitted field is

$$E_{out}(t) = Q(t) \int dt'' R(t - t'') y(t'') \tag{39}$$

The two-time correlation function of the filtered light is:

$$\left\langle E^*_{out}(t + \tau) E_{out}(t) \right\rangle = N_y Q^*(t + \tau) Q(t) \int dt' R(t + \tau - t') R(t - t') \tag{40}$$

showing that for a sufficiently narrow spectral width, the pulse amplitude goes to zero before the coherence ( $\int dt' R(t + \tau - t') R(t - t')$ ) decays.

Before proceeding to specific examples, we wish to give 'universal' definitions of the 'bandwidth' $B$ of an incoherent spectral filter and the 'duration' $T$ of an incoherent temporal filter. To simplify the discussion, we state that both filters have unity value at their respective peaks, $\tilde{R}(\omega = 0) = 1$, $Q(t = 0) = 1$. If this is not case, an overall multiplicative factor ( $\tau_F$ ) can be applied to the product $| \tilde{R}(\omega) \tilde{Q}(\omega - \omega') |^2$, to account for the frequency-and-time-independent losses. It can be absorbed into an overall channel or 'insertion' loss of the filter device. With this assumption, we are motivated by Eq.(7), which for a SIF reads, from Eq.(35) , to define a time-bandwidth product by:

$$\sum_{n=0}^{\infty} \lambda_n^2 = \int d\!\!\!/\,\omega \, | \tilde{R}(\omega) |^2 \int d\!\!\!/\,\omega' \, | \tilde{Q}(\omega - \omega') |^2 \tag{41}$$

$$\equiv BT$$

where, using Parseval's theorem (and recalling $d\!\!\!/\,\omega = d\omega / 2\pi$ ),

$$\int d\!\!\!/\,\omega \, | \tilde{R}(\omega) |^2 = \int dt \, | R(t) |^2 \equiv B$$
$$\int d\!\!\!/\,\omega' \, | \tilde{Q}(\omega - \omega') |^2 = \int dt \, | Q(t) |^2 \equiv T \tag{42}$$

$T$ is the duration in units of seconds. The bandwidth in units of $Hz$ is $B$, whereas in units of rad/s it is $2\pi B$. Such an integral is a good measure of filter width because unit transmission is in principle possible over an arbitrarily wide range of frequency or time. The same definition of bandwidth was used in a study of spectral filtering of quantum light. [26] Here we extend it to variable temporal filtering as well. We will see in the examples below that these definitions characterize well the bandwidth and the duration of the filters. Equation (41) has the satisfactory interpretation that the time-bandwidth product gives the effective number of modes passed by the SIF. It also allows us to write a compact, intuitive expression for the discrimitivity, from Eq.(12):

$$\xi = \frac{\eta}{BT} \tag{43}$$

## 6.  Coherent TF filters

In contrast to a SIF, a coherent TF filter can, in principle, create a perfectly coherent pulse when a stationary, broadband incoherent field $y(t)$ passes through it. If, as in Eq.(11) , one of the singular values equals one, with all others zero, we will obtain:

$$E_{out}(t) = \lambda_0 \psi_0(t) \int dt' \varphi_0^*(t') y(t') \tag{44}$$

where the integral selects (projects) that portion of $y(t)$ that occupies the input TM $\varphi_0(t)$. Here, in this ideal limit, which is in-principle reachable, $E_{out}(t)$ is perfectly coherent, although it has overall random phase and amplitude, determined by the particular realization of $y(t)$ during the filtering process.

In this sense, coherent TF filtering of incoherent light created short pulses having properties nearly identical to stimulated Raman pulses created in molecular systems driven by an ultrashort laser pulse. There the scattering process acts like a coherent TF filter for the quantum vacuum fluctuations that spontaneously seed the Raman build-up process. [20]

## 7.  Example – Gaussian SIF

The question is - how close to the optimum that is provided by the ideal coherent TF filter can be achieved by a (two-stage) sequential incoherent TF filter? Consider a SIF where both stages are Gaussian functions in frequency (and in time):

$$\tilde{R}(\omega) = \exp(-\omega^2 \pi / 2[2\pi B]^2)$$
$$\tilde{Q}(\omega) = T\sqrt{2\pi} \exp(-\omega^2 T^2 \pi / 2) \tag{45}$$

where $B$ characterizes the frequency filter's bandwidth (in Hz) and $T$ characterizes the time filter's opening time, consistent with the definitions in Eq. (42).

Note here we approximate the kernel as one that does not equal zero for $t' > t$. To retain causality in this case, we need to examine the output pulse long after it fully exists the filter device, so that the kernel is vanishingly small in this time range. In such cases it is common to shift the output time variable to a local time, so the pulse appears near the zero time in the shifted variable.

If the frequency filter acts first, we have:

$$\tilde{F}_{TF}(\omega, \omega'') = T\sqrt{2\pi} \exp(-(\omega - \omega'')^2 \pi T^2 / 2) \exp(-\omega''^2 \pi / 2[2\pi B]^2) \tag{46}$$

The two-dimensional Gaussian is one of the few that has a known analytical Schmidt decomposition, known as the Mehler identity [27, 28, 29, 30]

$$\exp\left(-\frac{1-u}{1+u}\frac{(x+y)^2}{4} - \frac{1+u}{1-u}\frac{(x-y)^2}{4}\right) = \sum_{n=0}^{\infty} \sqrt{\pi(1-u^2)}\, u^n\, HG_n(x)HG_n(y) \tag{47}$$

valid for $-1 < u < 1$, where $HG_n(x)$ are the square-normalized Hermite-Gaussian functions: $HG_n(x) = (\pi^{1/2} 2^n n!)^{-1/2} H_n(x) \exp(-x^2 / 2)$, $H_n(x)$ being the nth-order Hermite polynomial. For the Gaussian filter, we have (after considerable algebra):

$$u = \sqrt{1 + (2BT)^{-2}} - (2BT)^{-1} \tag{48}$$

and then:

$$F_{TF}(\omega, \omega') = \sum_{n=0}^{\infty} \lambda_n \sqrt{2\pi / \beta} \, HG_n(\omega / \beta) \sqrt{2\pi / \alpha} \, HG_n(\omega' / \alpha) \qquad (49)$$

where the frequency variable scalings are

$$\alpha = \frac{\sqrt{\pi}}{T} \left(1 + (2BT)^2\right)^{1/4}$$

$$\beta = \frac{2\sqrt{\pi} B}{\left(1 + (2BT)^2\right)^{1/4}} \qquad (50)$$

and the singular values are:

$$\lambda_n = u^{n+1/2} \qquad (51)$$

It's worth mentioning that $T^2 \alpha \beta = 2BT$. Thus, the 'natural' input and output modes of the Gaussian SIF are Hermite-Gaussians, normalized as in Eq.(9), with different durations:

$$\tilde{\psi}_n(\omega) = \sqrt{2\pi / \beta} \, HG_n(\omega / \beta)$$

$$\tilde{\varphi}_n(\omega') = \sqrt{2\pi / \alpha} \, HG_n(\omega' / \alpha) \qquad (52)$$

The time-bandwidth product $BT$ is seen to be the single characteristic parameter for the Gaussian SIF.

The efficiency for transmitting the optimal (target) mode is $\eta = \lambda_0^2 = u = \sqrt{1 + (2BT)^{-2}} - (2BT)^{-1}$. The descrimitivity defined by Eq.(12) evaluates to $\xi = 1 - u^2 = 1 - \eta^2$.

In Fig. 2(a) the Gaussian filter singular values are plotted for two choices of $BT$. The distributions show that for each value of $BT$, the eigenfunction $\psi_0(t)$ is maximally concentrated in the time window defined by the temporal filter. They also indicate a tradeoff between descrimitivity (how rapidly the singular values drop off) and efficiency (the magnitude of the largest singular value).

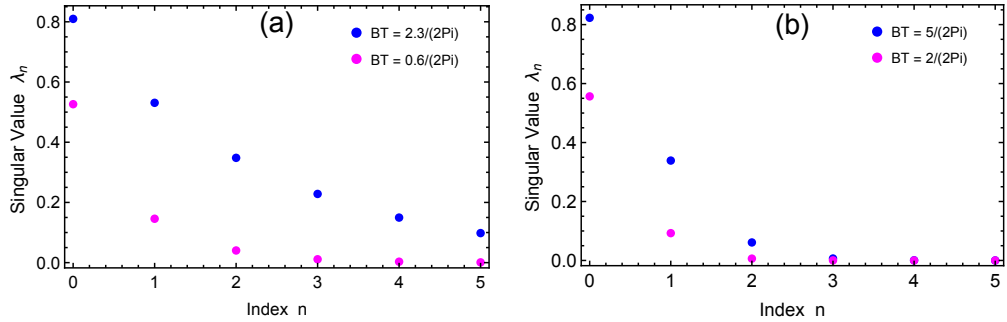

**Figure 2.** (a) Gaussian filter singular values versus mode index $n$ for two choices of $BT$: $2.3/2\pi$ and $0.6/2\pi$. (b) Rectangular filter singular values versus mode index n for two choices of $BT$: $5/2\pi$ and $2/2\pi$ [See Sec. 8]. $BT$ values are chosen so singular values are comparable for mode index $n = 0$ in the two plots.

In Fig.3(a) we plot the efficiency-descrimitivity tradeoff for the Gaussian SIF. We see for this filter type a fairly rapid fall off of descrimitivity with increasing efficiency.

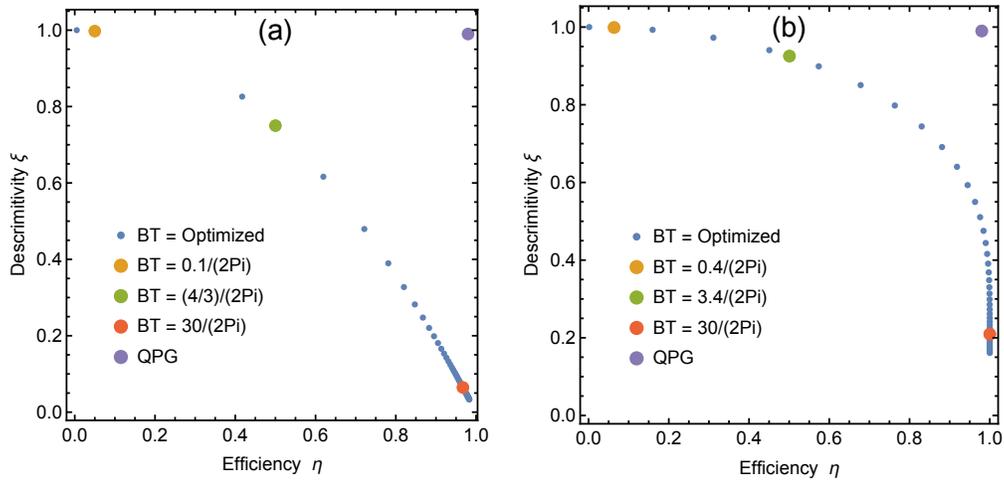

**Figure 3.** (a) Efficiency-descrimitivity tradeoff for Gaussian SIF, wherein $BT$ varies from $0.01/2\pi$ to $60/2\pi$. (b) Same for rectangular (Slepian) SIF of Sec. 8, wherein $BT$ varies from $0.01/2\pi$ to $50/2\pi$; the point at $(.99, .98)$ is the result for a coherent TF filter as discussed in Section 9. The point in the upper-right corner of each panel is for the two-stage QPG, predicted to provide target-mode efficiency $\eta = 0.99$ with mode descrimitivity $\xi = 0.98$. [9]

Note that for the Gaussian SIF, the time filter is Hermitian, $\tilde{Q}^*(\omega'-\omega) = \tilde{Q}(\omega-\omega')$, and the frequency filter is real, $\tilde{R}^*(\omega) = \tilde{R}(\omega)$, therefore the singular values are independent of the filter order and the forms of the input and output modes are simply swapped.

As an example of a coherent TF filter that outperforms the sequential Gaussian SIF, we include a point in the upper-right corner of Fig. 2(a) for the quantum pulse gate (QPG). The most-studied example of the QPG is based on sum-frequency generation by three-wave mixing in a nonlinear optical crystal. [5, 7, 15, 10] It was predicted and demonstrated that improved performance of this type of QPG is obtained using a two-stage double-pass configuration, [8, 10] The two-stage QPG is predicted to provide target-mode efficiency $\eta = 0.99$ with mode descrimitivity $\xi = 0.98$. [9].

## 8.  Example – rectangular (Slepian) SIF

Consider a SIF where both stages are 'rectangular' or 'top hat' functions, defined consistent with the normalization conditions Eq.(42). The frequency filter is:

$$\tilde{R}(\omega) = \begin{cases} 1 & for\ |\omega| < \Omega \\ 0 & otherwise \end{cases} \tag{53}$$

$$R(t) = (\Omega / \pi)\operatorname{sinc}(\Omega t)$$

where $\Omega = \pi B$ and $\operatorname{sinc}(x) \equiv \sin(x)/x$. The input modes are bandlimited to the (angular frequency) domain $[-\Omega, \Omega]$, i.e. the frequency domain, in Hz, $[-B/2, B/2]$ The temporal filter is:

$$Q(t) = \begin{cases} 1 & for\ |t| < \tau \\ 0 & otherwise \end{cases} \tag{54}$$

$$\tilde{Q}(\omega) = 2\tau \operatorname{sinc}(\omega\tau)$$

where $\tau = T/2$. Therefore, the output modes are time-limited to $[-\tau, \tau]$. The sequential filter kernel is then:

$$F_{FT}(t,t'') = Q(t'')\frac{\Omega}{\pi}\operatorname{sinc}(\Omega(t-t'')) \tag{55}$$

Equation (6) implies that the modes of this filter obey the integral equation:

$$\int_{-\tau}^{\tau} \frac{\sin(\Omega(t-t'))}{\pi(t-t')}\varphi_n(t')dt = \lambda_n\psi_n(t) \tag{56}$$

The input modes are bandlimited to (angular frequency range) $\omega \in [-\Omega, \Omega]$ and the output modes are time-limited to $t \in [-\tau, \tau]$. Thus, they must be normalized as:

$$\int_{-\infty}^{\infty} dt\, \psi^*_m(t)\psi_n(t) = \delta_{mn} \quad , \quad \int_{-\tau}^{\tau} dt\, \varphi^*_m(t)\varphi_n(t) = \delta_{mn} \tag{57}$$

By changing variable to $x = t/\tau$ one can show that the solutions depend only on the parameter $c = \Omega\tau = (\pi/2)BT$.

In a famous paper on signal processing, Slepian and Pollak ([21] referred to hereafter as S&P) discussed an integral eigenvalue equation of a form nearly identical to Eq.(56):

$$\int_{-\tau}^{\tau} \frac{\sin(\Omega(t-t'))}{\pi(t-t')} \Phi_n(t')dt = \beta_n \Phi_n(t) \tag{58}$$

The difference between Eqs.(56) and (58) is only in the domains of the solutions and their normalizations, but this difference is crucial to obtaining the correct singular values and temporal modes for the filter problem. The 'Slepian functions' remarkably satisfy two normalization conditions:

$$\int_{-\infty}^{\infty} dt\, \Phi_m(t)\Phi_n(t) = \delta_{mn} \quad , \quad \int_{-\tau}^{\tau} dt\, \Phi_m(t)\Phi_n(t) = \beta_n \delta_{mn} \tag{59}$$

We can compare the normalization conditions to find the relations between the filter modes and the Slepian functions:

$$\int_{-\infty}^{\infty} dt\, \Phi_m(t)\Phi_n(t) = \delta_{mn} \quad , \quad \int_{-\tau}^{\tau} dt\, \frac{\Phi_m(t)}{\sqrt{\beta_m}}\frac{\Phi_n(t)}{\sqrt{\beta_n}} = \delta_{mn} \tag{60}$$

So, the filter modes are:

$$\psi_n(t) \equiv \frac{\Phi_n(t)}{\sqrt{\beta_n}} \quad , \quad \varphi_n(t) = \Phi_n(t) \tag{61}$$

Then the integral equation (58) is:

$$\int_{-\tau}^{\tau} \frac{\sin(\Omega(t-t'))}{\pi(t-t')} \psi_n(t')dt = \sqrt{\beta_n}\varphi_n(t) \tag{62}$$

So, we see that the singular values are $\lambda_n = \sqrt{\beta_n}$.

S&P first pointed out that the solutions of (58) are angular prolate spheroidal functions of zero order, $S_{0n}(c,x)$, parametrized by $c$. The eigenfunctions in S&P are defined such that $\Phi_n(x)/\sqrt{\lambda_n}$ is square-normalized to unity in the interval $x \in [-1,1]$:

$$\Phi_n(t) = \sqrt{\frac{\beta_n}{\sqrt{\int\limits_{-1}^{1}\left[S_{0n}(c,x)\right]^2 dx}}} \, S_{0n}(c,t/\tau) \tag{63}$$

and are called *prolate spheroidal wave functions* (PSWFs), or sometimes *Slepian functions*. S&P showed that the PSWF have the following properties: *1.* They are strictly bandlimited to the spectral domain $[-\Omega, \Omega]$, *2.* They are orthogonal and complete for functions bandlimited to $[-\Omega, \Omega]$, and *3.* $\Phi_0(t)$ is the function that is maximally concentrated in time given that it is bandlimited to $[-\Omega, \Omega]$.

The eigenvalues are given by a second form of prolate spheroidal functions denoted $R_{0n}^{(1)}(c,x)$:

$$\lambda_n = \sqrt{\beta_n} = \sqrt{\frac{2c}{\pi}} R_{0n}^{(1)}(c,1) \tag{64}$$

Both needed forms of prolate spheroidal functions mentioned here can be evaluated using Mathematica, for example. (See the Appendix) In Fig. 2(b) the PSWF singular values are plotted for two choices of $BT$ $(i.e., 2c/\pi)$, chosen so the values for $n = 0$ match roughly those in Fig. 2(a) for the Gaussian filter, for which the SVs drop off more slowly. The distributions again show the tradeoff between descrimitivity and efficiency.

In Fig. 3(b) we plot the efficiency-descrimitivity tradeoff for the rectangular (Slepian) SIF. We see a far better performance for this filter type than for the Gaussian filter. Yet, if we wish to have both efficiency and descrimitivity greater than, say, 0.9, even this filter cannot achieve such performance.

To achieve the optimum behavior as shown in Fig. 3(b), we assume that we can create the input signal pulse exactly in the optimum mode of the filter, $\psi_0(t) \propto S_{00}(c,t/\tau)$. This function is plotted in Fig. 4 for several values of $c = (\pi/2)BT$. For larger values of c, the optimum mode begins to appear more Gaussian-like.

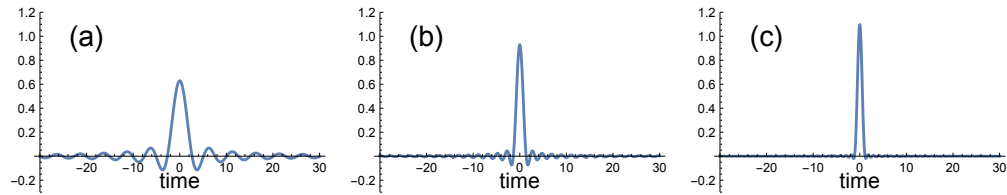

**Figure 4.** Optimum input mode $\psi_0(t)$ for rectangular SIF, for different values of $c = (\pi/2)BT$: (a) $c = 1.25$, (b) $c = 3.0$, (c) $c = 5.0$.

Because the input modes of the SIF are band limited, the output modes are simply related to the input modes by a truncation in time and an amplitude scaling to maintain normalization.

As with the Gaussian SIF, in this case the time filter is real and symmetric and the frequency filter is real, therefore if the order of time and frequency filtering are swapped, the singular values are unchanged and the forms of the input and output modes are swapped.

## 9. Example: entanglement-based quantum key distribution

As an exemplary application of the formalism developed here, consider generation of a cryptographic key using the BBM92 protocol [31] by detecting polarization-entangled photon pairs distributed in the presence of background noise. For simplicity we assume that the pairs emitted by the source are described by the maximally entangled two-qubit pure state $|\psi_+\rangle = (|HV\rangle + |VH\rangle)/\sqrt{2}$ and that the photons are generated in single, identical, well-defined temporal modes. Such pairs can be produced in a spontaneous parametric down-conversion process with a carefully chosen combination of the medium phase-matching function and the pump spectral width [32].

Suppose that the power transmission coefficient of the optical channel from the source to one of the receivers is $\tau \ll 1$ and that the background noise power spectral density expressed in photon number units is $N_y$. If a SIF is used, the mean number of background photons received in one slot is given by $N_y \eta / \xi$, from Eq.(16), which is assumed to be much less than one. On the other hand, the probability that a signal photon passes the SIF is $\eta$. Hence the overall probability of a coincidence detection event is $p_c = (\tau\eta + N_y\eta/\xi)^2$. Out of that, $(\tau\eta)^2$ is the probability that a genuine entangled pair is detected, while the rest is contributed by detection of two background photons or a combination of a background photon and one photon from a pair. Therefore, the normalized effective two-photon state detected by the receivers is of the form:

$$\rho = \frac{1}{(1 + n_y/\xi)^2}|\psi_+\rangle\langle\psi_+| + \left(1 - \frac{1}{(1 + n_y/\xi)^2}\right)\rho_{mix} \tag{65}$$

where $n_y = N_y/\tau$ is the scaled noise strength, and $\rho_{mix}$ is the normalized completely mixed two-qubit state. The quantum bit error rate QBER is equal to half of the contribution from the completely mixed state, and hence is given by:

$$QBER = \frac{1}{2}\left(1 - \frac{1}{(1 + n_y/\xi)^2}\right) \tag{66}$$

According to the standard analysis of the BBM92 protocol [33, 3], the attainable key rate is:

$$\begin{aligned}
R_K &= R_S p_c \left[1 - 2H(QBER)\right] \\
&= R_S(\tau\eta)^2(1 + n_y/\xi)^2\left[1 - 2H(QBER)\right]
\end{aligned} \tag{67}$$

where $R_S$ is the rate of photon pairs emitted by the source and $H(x) = -x\log_2(x) - (1-x)\log_2(1-x)$ is the binary entropy. The characteristics of the SIF enters through the factor $R_K/(R_S\tau^2) = \eta^2(1 + n_y/\xi)^2\left[1 - 2H(QBER)\right]$, where properties of the optical channel appear solely in the form of the ratio $n_y = N_y / \tau$. In Fig. 5(a) we plot $R_K / (R_S \tau^2)$ as a function of the SIF efficiency $\eta$, assuming that the descrimitivity achieves that of a Gaussian filter, i.e. $\xi = 1 - \eta^2$. It is seen that with increasing impact of the background noise strength, the filter efficiency has to be lowered in order to ensure that the transmitted noise is sufficiently suppressed, thereby reducing the key rate. In Fig. 5(b) we plot $R_K / (R_S \tau^2)$ as a function of the SIF efficiency $\eta$, assuming the descrimitivity achieves that of a Slepian filter, determined as in Sec. 8.

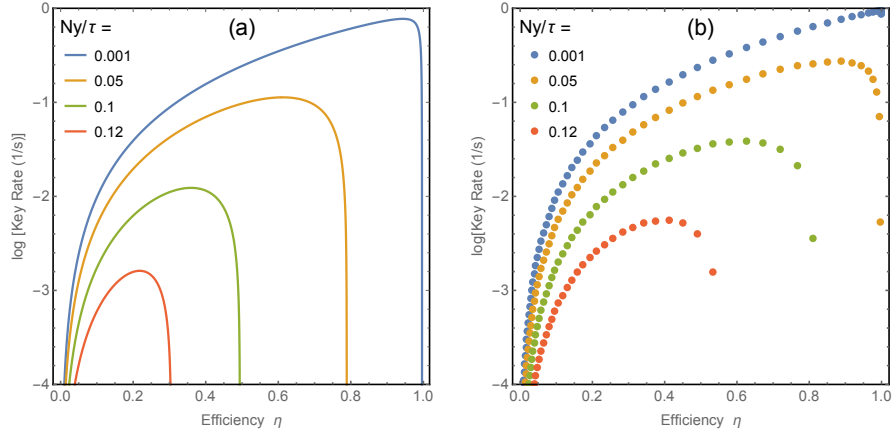

**Figure 5.** QKD rates (base-10 log scale) versus efficiency and various values of scaled background noise $N_y / \tau$, for (a) Gaussian SIF, (b) Slepian SIF.

The results with the Gaussian or Slepian SIF are in stark contrast to that using the quantum pulse gate, where it is possible to achieve simultaneously high efficiency and high descrimitivity. Note that in order to preserve the polarization entanglement, the quantum pulse gate would need to be applied to both polarization components of the incoming signal in a coherent fashion, and would need to be synchronized with the signal generated by the photon pair source. In order to illustrate the advantage of using the quantum pulse gate, in Fig. 6 we plot the maximum achievable normalized key rate, optimized to the peaks of the curves in Fig. 5, for four filter types: Gaussian SIF, Slepian SIF, 'practical coherent filter' (two-stage quantum pulse gate and 'perfect' coherent filter (the theoretical best possible TF filter, defined by Eq.(11) and reviewed in Sec. 6). It is seen that coherent filtering significantly outperforms the sequential incoherent filters, until the noise strength becomes so high that no kind or amount of filtering can yield a nonnegligible key rate.

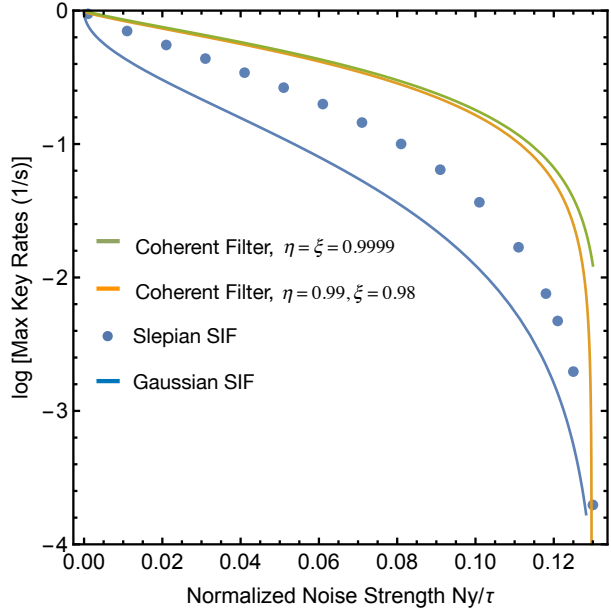

**Figure 6.** Maximum normalized key rate (base-10 log scale) with optimized efficiency, versus normalized noise strength, for four filter types: Gaussian SIF, Slepian SIF, 'practical coherent filter,' ( $\eta = 0.99, \xi = 0.98$ ), and 'perfect' coherent filter ( $\eta = \xi = 0.9999$ ).

## 10. Summary and conclusions

In this paper we presented a general theory of incoherent time-frequency filters for optical fields constructed as a sequence of an incoherent (time-stationary) frequency filter followed by a frequency-neutral temporal filter. A powerful mathematical tool to analyze properties of such filters is Schmidt decomposition, which allows one to identify the set of mutually orthogonal input temporal modes that are mapped one-to-one onto output temporal modes. The singular values of the decomposition specify the amplitude transmission coefficients between individual pairs of input and output modes. This framework enables characterization of a sequential incoherent filter (SIF) in terms of its efficiency $\eta$, which specifies the signal efficiency, i.e. transmitted fraction of the signal energy, and mode descrimitivity $\xi$, which describes the capability of the filter to discriminate and reject background noise, and thus to increase the signal-to-noise ratio.

We found a remarkably compact and intuitive expression relating efficiency and descrimitivity of any SIF, Eq.(43):

$$\xi = \frac{\eta}{BT} \tag{68}$$

where $B$ is the spectral filter's bandwidth in Hz, and $T$ is the temporal filter's duration in seconds, given by the 'universal' definitions in terms of the intensity transmission functions:

$$\int |\tilde{R}(\omega)|^2 \frac{d\omega}{2\pi} \equiv B \quad , \quad \int |Q(t)|^2 dt \equiv T \tag{69}$$

The time-bandwidth product $BT$ gives the effective number of modes passed by the SIF, and is independent of the detailed shapes of the filter functions. The signal its efficiency $\eta$ does depend on the shapes of the filter curves, as it is the square of the largest singular value of the filter function's decomposition.

We quantitatively explored two examples – Gaussian and rectangular (Slepian) SIFs – and presented the inherent trade-off between efficiency and descrimitivity in these cases. In the Gaussian case, the trade-off admits an elementary analytical form. The results are in strong contrast with properties of a coherent filter in the form of a quantum pulse gate, which in principle permits both efficiency and descrimitivity to approach simultaneously one.

We quantify the degree to which coherent filters can substantially improve the performance of quantum communication over noisy channels. As an example, we studied quantum key distribution, wherein strong filtering of background noise is necessary to maintain a high quality of entanglement producing non-classical correlations between detection events, while high signal transmission is of paramount importance to ensure a useful key generation rate.

The formalism presented can be extended to study a wide variety of combinations of time- and frequency filters as a way to optimize the characteristics of sequential incoherent filters for specific applications. It should be noted that in some scenarios, a more detailed study of the transformation of temporal modes may be necessary. For example, applying too strong frequency filtering to pulse position modulation signal may spread individual pulses outside the width of individual temporal slots constituting symbol frames and consequently introduce errors in symbol identification. Another important point is that the temporal characteristics of the signal should be carefully matched to the natural (Schmidt) modes of the time-frequency filter to ensure optimal operation. This may lead to interesting questions regarding the design of optical signal sources, especially in the case of non-classical light.

## Acknowledgements

We thank Brian Smith for useful discussions. MR was supported by NSF grants PHY-1820789 and QII-TAQS 1936321. KB was supported by the project "Quantum Optical Technologies" carried out within the International Research Agendas programme of the Foundation for Polish Science co-financed by the European Union under the European Regional Development Fund. This work was carried out in the context of the NSF Center for Quantum Networks, award number 1941583.

## Disclosures

The authors declare no conflicts of interest.

## Appendix

For convenience, we give the forms of Slepian singular values and functions as expressed in Mathematica:

$$\beta\big[n\_,c\_\big] := \frac{2c}{\pi}\Big(SpheroidalS1\big[n,0,BT/4,1\big]\Big)^2$$
$$\Phi\big[n\_,c\_,x\_\big] := N_0\ SpheroidalPS\big[n,0,c,x\big] \tag{70}$$

where $N_0$ is a normalization constant.